\documentclass[superscriptaddress,aps,prl,twocolumn,showpacs,amsmath,amssymb]{revtex4}

\usepackage{graphicx}
\usepackage{dcolumn}
\usepackage{bm}
\usepackage{times}

\begin{document}



\title{Fano Resonance Between Mie and Bragg Scattering in Photonic Crystals}
\author{M. V. Rybin}
\affiliation{Ioffe Physico-Technical Institute, St.Petersburg 194021, Russia}
\author{A. B. Khanikaev}
\affiliation{Toyohashi University of Technology, Toyohashi, Aichi 441-8580, Japan}
\affiliation{MQ Photonics Research Centre, CUDOS and Dept of Physics, Macquarie University, NSW 2109, Australia}
\author{M. Inoue}
\affiliation{Toyohashi University of Technology, Toyohashi, Aichi 441-8580, Japan}
\author{K. B. Samusev}
\affiliation{Ioffe Physico-Technical Institute, St.Petersburg 194021, Russia}
\author{M. J. Steel}
\affiliation{MQ Photonics Research Centre, CUDOS and Dept of Physics, Macquarie University, NSW 2109, Australia}
\author{G. Yushin}
\affiliation{Georgia Institute of Technology, Dept. Materials Science, Atlanta, GA 30332-0245, USA}
\author{M. F. Limonov}
\affiliation{Ioffe Physico-Technical Institute, St.Petersburg 194021, Russia}
\affiliation{Toyohashi University of Technology, Toyohashi, Aichi 441-8580, Japan}


\date{\today}
               
\begin{abstract}
We report the observation of a Fano resonance between continuum Mie scattering and a narrow Bragg band in synthetic opal photonic crystals. The resonance leads to a transmission spectrum exhibiting a Bragg dip with an asymmetric profile, which can be tunably reversed to a Bragg rise. The Fano asymmetry parameter is linked with the dielectric contrast between the permittivity of the filler and the specific value determined by the opal matrix. The existence of the Fano resonance is directly related to disorder due to non-uniformity of \textit{a}-SiO$_{2}$ opal spheres. Proposed theoretical ``quasi-3D'' model produces results in excellent agreement with the experimental data.
\end{abstract}

\pacs{42.70Qs, 42.25.Fx, 42.79.Fm}

\maketitle  

Mie and Bragg scattering are key optical phenomena in photonic crystals (PhC) composed of spherical or nearly spherical particles. Light scattering by an isolated spherical particle can be described by Mie theory \cite{1}. Considering PhC composed of a periodic array of such spheres, interference of scattered waves results in the transformation of Mie scattering into Bragg scattering and gives rise to formation of the photonic band structure \cite{2}. The underlying Mie scattering is therefore hidden in perfectly ordered PhC and as a result, it has been insufficiently studied so that its role is clear only for the case of perfect structures. The resulting Bragg scattering, on the other hand, has been intensively examined, both experimentally and theoretically in great detail \cite{2,3,4,5,6,7}. In particular, the multiple Bragg diffraction phenomenon in PhC, which occurs when two narrow Bragg bands demonstrate the avoided crossing effect \cite{7,8,9,10} has been thoroughly studied. Departing from this phenomenon and with intention to further deepen our understanding of light scattering in PhC, a number of challenging problems can be formulated: What can we expect if a spectrally narrow Bragg band interacts with a broad spectrum originating from certain scattering mechanisms such as Mie or Fabry-Perot scattering? Is it possible to observe the consequences of this interaction or simply Mie scattering experimentally? What are the effects of inherent disorder in the structural components of opal-based PhC, beyond the well-known broadening and degradation of stop bands \cite{6,11,12,13,14,15}?

If a narrow Bragg band interacts with the continuum spectrum through an interference effect constructively or destructively, we can expect an interaction of Fano-type \cite{16}, a phenomenon well-known across many different branches of physics. The Fano resonance between continuum and discrete states manifests as an asymmetric profile of narrow band in the transmission spectrum, which in general has the form:
\begin{equation}
F(\Omega)=\frac{(\Omega+q)^2}{\Omega^2+1}
\label{eqn:Fano}
\end{equation}
where $\Omega=(\omega-\omega_{B})/(\gamma_{B}/2)$, $\omega_{B}$ is the frequency, $\gamma_{B}$ is the width of the narrow band, and $q$ is the Fano asymmetry parameter. It was shown theoretically that Fano-type asymmetric line shapes can be created in the response function of certain PhC. In general these systems consist of a waveguide with forward and backward propagating waves being indirectly coupled via one or more mediating resonant cavities, defects or another similar mechanism \cite{17,18,19,20,21,22,23}. Surprisingly, despite this growing interest in theoretical and experimental studies of Fano resonances in PhC, until now no one has ever considered a Bragg band acting as the discrete narrow band, as opposed to resonant cavities or defect states. 

In this Letter, we report on the experimental discovery of \emph{Fano resonances involving interference between Mie scattering (as a continuum) and Bragg scattering (as a discrete state)}. The (111) Bragg band in the transmission spectra of synthetic opals demonstrates all characteristic features of the Fano resonance, including asymmetry of the transmission band which in this case can be inverted around its central frequency when the sign of the Fano parameter q is reversed. Moreover, we report observation of an impressive effect of the Bragg band reversal, which manifests as appearance of a transmission ``rise'' instead of a transmission dip, and which we refer to as ``\emph{enhanced Bragg transmission}''. We also propose a theoretical ``quasi-3D'' model able to reproduce all specific features of the experimentally observed Fano resonance.

To study the general features of light scattering in PhCs, we chose synthetic opals as an archetypal 3D PhC \cite{6,7,10,11,12,24,25,26}. An fcc lattice of opals is formed from nearly spherical \textit{a}-SiO$_{2}$  particles which are neither uniform in size nor homogeneous in dielectric permittivity \cite{11,26,27}. To obtain precise information about the inherent inhomogeneity of the \textit{a}-SiO$_{2}$ particles and statistical characteristics of the ensemble of particles in the investigated samples, scanning electron microscopy (SEM) micrographs were analyzed using an original 
\begin{figure}[!ht]
\centering
\includegraphics[width=0.45\textwidth]{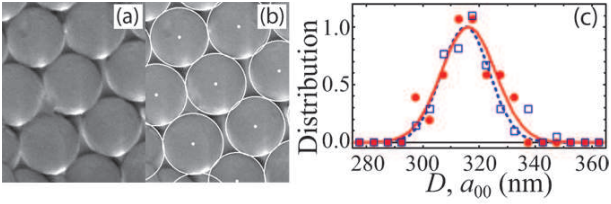}
\caption{(Color online) (a) The SEM image of the (111) opal plane with \textit{a}-SiO$_{2}$ particles. (b) The circles and the centers of circles derived by processing the SEM image  (a) using Hough-type procedure. (c) The distributions of the circle diameters $D$ (red circles) and the distances between circle centres $a_{00}$ (blue squares) obtained by processing the image (b). The curves are the results of fitting with a Gaussian function.}
\label{fig1}
\end{figure}
procedure based on the Hough transform \cite{28}.
From the micrograph shown in Fig.1(b) we obtain several important structural parameters of the sample under study, such as the average diameter of the \textit{a}-SiO$_{2}$ particles $\bar{D}$=316.2nm and full-width at half-maximum of its normal (Gaussian) distribution of 22.8 nm. The latter corresponds to deviation in the diameter $D$ equal to 7.2\% at half maximum. The average distance between particle centres $\bar{a}_{00}$=315.2 nm, while  $\bar{D}>\bar{a}_{00}$  because of sintering of neighbouring \textit{a}-SiO$_{2}$ spheres in opals. 

We investigated transmission spectra of two samples of \textit{a}-SiO$_{2}$ particles with average diameters of 260 nm and 316 nm respectively. The sample was placed at the center of a vessel with plane-parallel quartz windows, which was filled with an immersion liquid. The liquid served as the opal filler and as the ambient medium for the sample. The permittivities
of the liquids were precisely measured by an Abbe IRF-454B2M refractometer \cite{11,26}. Two liquids - distilled water with $\varepsilon_{f}$=1.778 and propylene glycol with $\varepsilon_{f}$=2.053, as well as their mixtures were used, providing variation in the dielectric constant of the filler in the range of $1.778\leq\varepsilon_{f}\leq2.053$. 

The transmission spectra of synthetic opals were recorded by a Perkin-Elmer Lambda-650 spectrophotometer in the 215-850 nm wavelength range when white light was passed through an oriented opal sample along the $\left[111\right]$-axis. In this scattering geometry, the (111) stop band is dominant (Fig. 2), and its spectral position is determined by $\lambda_{(111)}\approx\sqrt{8/3}\bar{D}\sqrt{g\bar{\varepsilon_{s}}+(1-g)\varepsilon_{f}}$, where $g\approx0.74$ is the fill fraction of the \textit{a}-SiO$_{2}$ particles which have an average dielectric permittivity $\bar{\varepsilon_{s}}$=1.92 \cite{11}. Careful measurements of the real $\varepsilon_s(r)$ profile function remains a challenge due to the small dimensions of the spheres. A possible information source might be the particle vibration modes \cite{TStill}. Higher Miller-index ($hkl$) stop bands are also seen in the 215-450 nm wavelength range [11]. Besides all these stop bands, the transmission spectrum features a continuum broad part with an abrupt decrease in the transmission at about $\lambda<{\lambda_{(111)}}/2$. This specific shape was reproduced in a number of works on spectroscopy of PhC composed of spherical particles and therefore can be attributed as a feature typical of these structures \cite{5,10,11,12,24,25,26}.
\begin{figure}[!ht]
\centering
\includegraphics[width=0.45\textwidth]{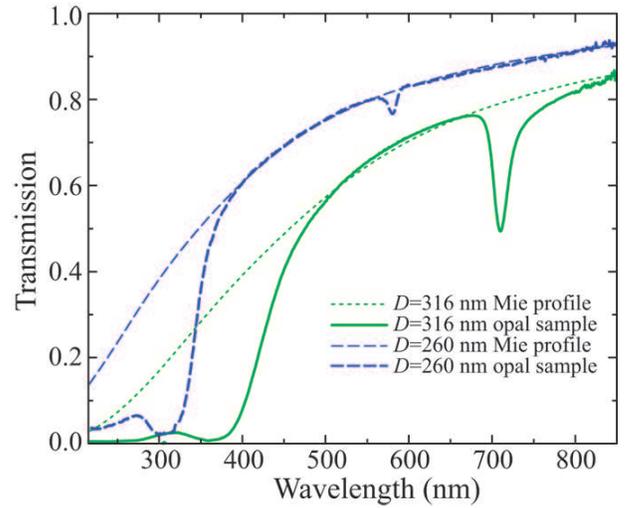}
\caption{(Color online) Transmission spectra of two opal samples ($D$=260 nm, and $D$=316 nm) for the $\Gamma\rightarrow L$ scattering geometry and the results of calculations of the transmission spectra of an disordered ensemble of isolated spherical particles with $D$=260 nm and $D$=316 nm with use of Mie theory.}
\label{fig2}
\end{figure}
Figure 2 shows a sharp and clear (111) band in the transmission spectrum, and this fact allowed us to analyze its shape especially thoroughly. The results clearly indicate a remarkable transformation of the (111) band when the filler permittivity $\varepsilon_{f}$ is changed [Fig. 3(a)]. At first, the well-known effect of diminishing of the stop band intensity is observed when the filler permittivity   approaches a certain value, which was determined for the family of (111) stop bands of the 316 nm opal sample to be $\varepsilon^{0}_{f}(111)=1.816$. The dependence of this value on ($hkl$) index of stop band family is due to the inhomogeneity of the \textit{a}-SiO$_{2}$ particles \cite{11,26}. In this work, however, we focus on a different phenomenon which can be clearly observed in the transmission spectra, but has not, to our knowledge, previously been discussed. We observe an asymmetry of the (111) dip which is inverted around a central frequency when the sign of the dielectric contrast $\varepsilon_{f}-\varepsilon^{0}_{f}$ is reversed. For $(\varepsilon_{f}-\varepsilon^{0}_{f})<0$ (green curves in Fig. 3), the long-wavelength wing of the (111) stop band is relatively flat in contrast to a steep short-wavelength wing. For $(\varepsilon_{f}-\varepsilon^{0}_{f})>0$ (blue curves in Fig. 3), the situation reverses - the long-wavelength wing of the (111) stop band becomes abrupt as compared to the now relatively flat short-wavelength wing of the band. Quite surprisingly, we found that there is no value of $\varepsilon_{f}$ \emph{at which the (111) band completely disappears}. Moreover, when the filler permittivity is $\varepsilon_{f}=\varepsilon^{0}_{f}$ (red curves in Fig. 3), \emph{a transmission Bragg rise} is observed instead of the conventional transmission Bragg dip; \emph{i.e. we have discovered the ``enhanced Bragg transmission''}. Note that the transmission Bragg rise could only be observed in a very narrow range of the filler permittivity values $\Delta\varepsilon_{f}\approx 1.816\pm0.003$. Therefore, this phenomenon was overlooked in our previous papers (Refs.~\onlinecite{11,26}), where the filler permittivity was changed using a larger step ($\Delta\varepsilon_{f}\sim 0.01 –- 0.03$).
\begin{figure}[!ht]
\centering
\includegraphics[width=0.45\textwidth]{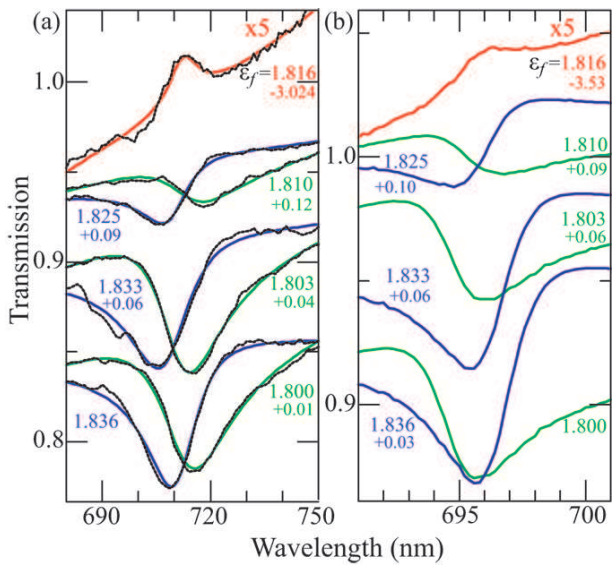}
\caption{(color) (a) The transmission spectra of an opal sample ($D$=316 nm, thickness $\approx$ 0.6 mm) as a function of the filler permittivity $\varepsilon_{f}$ in the region of the (111) photonic band (black curves). The color curves (red, blue and green) are the results of fitting with a Fano function. (b) The results of calculations of the transmission spectra using the ``quasi-3D'' model of disordered opal structure described in the text: number of layers was taken to be 1000 and averaging was done over 40000 realizations for $\varepsilon_{f}$=1.816 and 10000 realizations for other values of $\varepsilon_{f}$. In (a) and (b) the curves are shifted vertically by the values shown. } 
\label{fig3}
\end{figure}

From a theoretical point of view, light scattering in PhC is inherently a multi-level process. The first level represents scattering by a single particle with its own scattering properties. At the higher levels, multiple scattering events by the identical particles arranged in a periodic structure add up, resulting in formation of the Bloch waves of a perfect PhC. Disorder, however, induces additional random phase shifts and amplitude variations at the first-level, single-scattering event. This, in turn, modifies higher level multiple scattering events and causes degradation of bands and Bloch modes, as well as giving rise to the appearance of broad background continuum. Finally, Bloch waves and the background continuum can interfere, giving rise to a Fano-type resonance.

It is now clear that in order to describe the experimentally-observed phenomenon, a theoretical model should incorporate both scattering mechanisms (Bragg and Mie), in a way that naturally permits their coupling. In the model we propose here, the scattering properties of a disordered array of spherical particles forming a two-dimensional (2D) layer, corresponding to the (111) plane of the fcc opal structure, are calculated at the first stage with use of the Mie theory \cite{1}. In-layer diffraction and multiple scattering processes are neglected in this model, which restricts the model applicability to the low contrast regime and implies that we neglect ordering of the particles within the layer. It can be shown that in this approximation, the transmittance and reflection of the layer are directly related to the forward and backward scattering amplitudes $S(\theta=0)$ and $S(\theta=180^{\circ})$ of Mie theory \cite{1}, respectively, through 
\begin{subequations}
\label{eq:second}
\begin{equation}
t_n=e^{ikh}\left(1-\frac{2{\pi}{\eta}}{k^2}\tilde{S_{n}}(\theta=0)\right),
\label{subeq:secondA}
\end{equation}
\begin{equation}
r_n=e^{ikh}\frac{2{\pi}{\eta}}{k^2}\tilde{S_{n}}(\theta=180^{\circ}).
\label{subeq:secondB}
\end{equation}
\end{subequations}
In the last expressions $n$ is the layer index, $k$ is the modulus of the wavevector in the filler material, $h$ is the layer thickness, $\eta$ is the density of the particles in the layer, and the tilde above the scattering amplitudes implies intralayer averaging over the particles' size $r$ and dielectric constant $\varepsilon$. This averaging is conducted for every single layer by numerical integration of the scattering amplitudes around average values $\tilde{\varepsilon}_n$ and $\tilde{r}_n$ which vary from layer to layer:
\begin{equation} 
\tilde{S}_n(\theta)=\int^{\infty}_{-\infty}\int^{\infty}_{-\infty}S(\theta,\varepsilon,r)f_n(\varepsilon-\tilde{\varepsilon}_n,r-\tilde{r}_n)d{\varepsilon}dr,
\label{eqn:Average}
\end{equation}
here $f_n(\varepsilon,r)$ is a probability density function taken in our calculations to be normal (Gaussian) distribution.

At the second stage, the optical properties of the fcc photonic crystal built from a sequence of such (111) layers are determined by the conventional 2x2 transfer matrix technique \cite{29}. The transfer matrix of every layer of the structure is expressed through transmission and reflection coefficients obtained at the previous stage with account of interlayer disorder, i.e., as was explained before, the permittivity of particles and their size also vary from layer to layer. Finally, the calculated transmittance is averaged over a sufficient number of realizations. This model therefore can be referred to as ``quasi-3D'' since the Bragg diffraction is purely 1D, while the disorder induced scattering is 3D (both intralayer and interlayer disorder are taken into account). In our calculations, the particle size distribution was taken to be normal (Gaussian) with full-width at half maxima of 7.2\% for average diameter of $\bar{D}$=316.2 nm (Fig.1). The same model was taken for modeling the distribution of permittivity. The only adjustable parameter was average sphere permittivity. As can be seen from calculated spectra shown in Figs.(2,3), all peculiarities of the experimental spectra, such as asymmetry of the Bragg dip, appearance of the rise corresponding to enhanced Bragg transmission, and the overall behavior of the transmission spectrum are revealed by our model. Additional calculations using other permittivity distributions $\varepsilon_s$ (such as rectangular or triangular), also demonstrated the presence of the Fano resonance, including the appearance of the Bragg rise.

We have now elucidated a scenario of light scattering in realistic PhC, which are always imperfect due to the presence of inherent disorder of their constitutive elements. The disorder breaks a condition of ideal Bragg scattering, which determines all properties of perfect structures, and gives rise to an additional scattering component. This defect-induced extra-scattering produces background radiation and is responsible for formation of a continuum spectrum with its specific characteristics determined by the light scattering by elements constituting PhC. In particular, in disordered 1D PhC composed of dielectric slabs, it would be extra Fabry-Perot scattering due to fluctuations in slabs permittivity, while in 3D PhC formed from spherical particles, this mechanism is Mie scattering. That is, it is the defect-induced Mie scattering which is responsible for and determines the character of the continuum spectrum of opals (Fig.2). The latter is characterized by the abrupt decrease in the transmission observed in the short wavelength part of the spectra  ${\lambda}<{{\lambda_{111}}/2}$ for various PhC formed from spherical particles \cite{5,10,11,12,24,25,26}. An increase in the scattering intensity responsible for this feature is a well-known from Mie theory \cite{1}. This is why transmission through collection of particles obtained within Mie theory for both $D$=260nm and $D$=316nm nicely reproduce this feature. The continuum defect-induced Mie spectrum interfere with narrow Bragg bands, and, according to our calculations, this effect is related, in the first place, with inhomogeneity of the \textit{a}-SiO$_{2}$ particles with respect to dielectric permittivity ${\varepsilon}_{s}$.
\begin{figure}[!ht]
\centering
\includegraphics[width=0.45\textwidth]{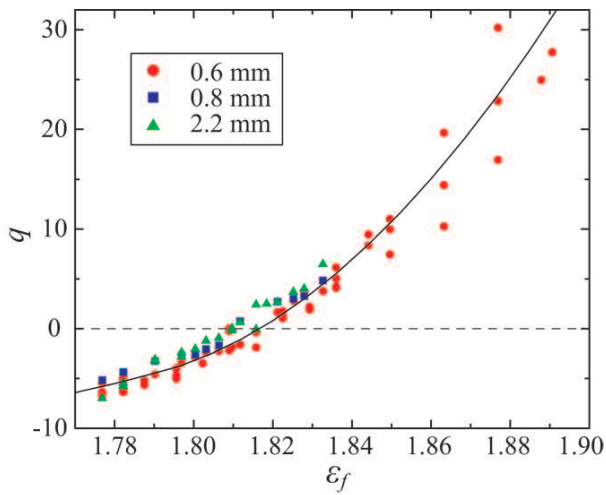}
\caption{(Color online) The Fano asymmetry parameter $q$ that changes continuously from negative to positive values with filler permittivity $\varepsilon_{f}$ increasing for three samples with different thickness of 0.6, 0.8, and 2.2 mm, $D$=316nm. The solid line is a guide for the eyes only.} 
\label{fig4}
\end{figure}
In addition we obtained good fitting of the experimental spectra by calculations with use of Eq. (1). This fact confirms that Fano-type interference between the continuum defect-induced Mie spectrum and narrow Bragg band takes place. Moreover, when Fano parameter $q$ sign is reversed [which corresponds to reversal in the sign of dielectric contrast (Fig. 4)], this results in inversion of (111) transmission dip (Fig. 3). The fact that sign reversal of parameter $q$ coincides with sign reversal of dielectric contrast $(\varepsilon_{f}-\varepsilon^{0}_{f})$ indicates that when $\varepsilon_{f}=\varepsilon^{0}_{f}$ reversal of the phase difference between disorder-induced background radiation and Bloch wave takes place. As it follows from analysis of curves obtained from Eq. (1) when $q$=0 narrow band should transform from transmission dip into transmission rise, which indeed is observed in both experimental and calculated spectra (red curves in Fig. 3).

To conclude, we report a beautiful example of a Fano-type resonance in PhC, which originates from interference between narrow Bragg band and the continuum spectrum. This distinctive continuum spectrum observed in opals is for the first time related with residual defect-induced Mie scattering. The ``quasi-3D'' theoretical model proposed for description of the light scattering in opals with account of disorder allowed us to depict all experimentally observed peculiarities. We found that the 3D nature of the disorder, especially in-plane positional disorder and from plane to plane variation of permittivity, is essential for explanation of the experimental results.

In a pioneering work \cite{30} on light scattering in disordered PCs it was predicted that localization of photons may occur in the vicinity of band edges. Since then such interplay between Bragg diffraction and disorder-induced scattering was considered only as a way to localize light and as a mechanism capable to suppress transmittance and increase back-scattering. Our studies, however, demonstrate that it is not always so and at some particular conditions Fano interference between scattered background radiation and Bloch waves gives rise to anomalous increase of the optical transmittance near photonic band edges or when Bragg condition is satisfied.

\acknowledgments{We thank A. A. Kaplyanskii and A. V. Baryshev for discussions; N. Stefanou and V. Yannopapas for discussions and useful references. This work was supported in parts by the RFBR, Russia, Grant 08-02-00642; CUDOS is an Australian Research Council Centre of Excellence. The work at TUT was supported in part by Grant-in-Aid for Scientific Research (S) No. 17106004 from Japan Society for the Promotion of Science (JSPS) and the Super Optical Information Memory Project from the Ministry of Education, Culture, Sports, Science and Technology of Japan (MEXT).}

\end{document}